\documentclass[12pt,preprint]{aastex}

\newcommand{\cer}{$E_{J-H}/E_{H-K_S}$}

\slugcomment{final version; 061221}

\shorttitle{Near-infrared extinction in Coalsack}
\shortauthors{Naoi et al.}

\begin{document}

\title{Near-infrared extinction in the Coalsack Globule 2}

\author{Takahiro Naoi\altaffilmark{1},
Motohide Tamura\altaffilmark{2},
Tetsuya Nagata\altaffilmark{3},
Yasushi Nakajima\altaffilmark{2},
Hiroshi Suto\altaffilmark{2},
Koji Murakawa\altaffilmark{4},
Ryo Kandori\altaffilmark{2},
Sho Sasaki\altaffilmark{5},
Shogo Nishiyama\altaffilmark{2},
Yumiko Oasa\altaffilmark{6},
and Koji Sugitani\altaffilmark{7}}

\altaffiltext{1}{ISAS, JAXA, 3-1-1 Yoshinodai, Sagamihara-shi, 
Kanagawa 229-8510, Japan}
\altaffiltext{2}{NAOJ, NINS, 2-21-1 Osawa, Mitaka-shi, Tokyo 181-8588, Japan}
\altaffiltext{3}{Department of Astronomy, Kyoto University, 
Kitashirakawa-Oiwake-cho, Sakyo-ku, Kyoto 606-8502, Japan}
\altaffiltext{4}{Max-Planck-Institut f\"ur Radioastronomie, 
Auf dem H\"ugel 69, D-53121 Bonn, Germany}
\altaffiltext{5}{NAOJ, NINS, 2-12 Hoshigaoka, Mizusawa-shi, 
Iwate 023-0861, Japan}
\altaffiltext{6}{Graduate School of Science and Technology, Kobe University, 
1-1 Rokkodai, Nada-ku, Kobe 657-8501, Japan}
\altaffiltext{7}{Graduate School of Natural Sciences, Nagoya City University, 
Mizuho-ku, Nagoya 467-8501, Japan}


\begin{abstract}

We have conducted $J$, $H$, and $K_S$ 
imaging observations for the Coalsack Globule 2 
with the SIRIUS infrared camera on the IRSF 1.4 m telescope at SAAO, 
and determined the color excess ratio, $E_{J-H}/E_{H-K_S}$. 
The ratio is determined in the same photometric system as 
our previous study for the $\rho$ Oph and Cha clouds 
without any color transformation; 
this enables us to directly compare 
the near-infrared extinction laws among these regions.
The current ratio $E_{J-H}/E_{H-K_{S}} = 1.91 \pm 0.01$ for 
the extinction range $0.5 < E_{J-H} <1.8$ 
is significantly larger than the ratios for the 
$\rho$ Oph and Cha clouds ($E_{J-H}/E_{H-K_{S}} = 1.60$--1.69).
This ratio corresponds to a large negative index $\alpha = 2.34 \pm 0.01$ 
when the wavelength dependence of extinction is approximated 
by a power law $A_\lambda \propto \lambda ^{-\alpha}$, 
which might indicate 
little growth of dust grains, or 
larger abundance of dielectric non-absorbing components 
such as silicates, or both 
in this cloud. 
We also confirm that 
the color excess ratio for the Coalsack Globule 2 has a trend 
of increasing with decreasing optical depth, 
which is the same trend as the $\rho$ Oph and Cha clouds have.  

\end{abstract}

\keywords{dust, extinction --- ISM: globules --- ISM: individual (Coalsack)}


\section{Introduction}

Precise determination of the wavelength dependence of extinction is 
necessary to understand the properties of interstellar dust 
because differences in the extinction law are attributed to
a variation in the dust grains. 
Differences in extinction laws among various lines of sight throughout the 
optical and UV properties of the spectrum are widely accepted 
(e.g., Fitzpatrick 1999). 
The existence of a corresponding variation at somewhat longer wavelengths 
is a controversial issue \citep{whi88,mat90,ken98}.
For near-infrared wavelengths, recent studies (e.g., Messineo et
al. 2005, Nishiyama et al. 2006) have found that the extinction law 
seems to change from one line of sight to another.
These results indicate that the so-called universality 
(e.g., Mathis 1990) does not necessarily hold for infrared wavelengths.  

The color excess ratio, \cer, is one of the simplest parameters available to 
express the near-infrared extinction law, which can be determined 
only from $J$, $H$, and $K_S$ photometric observations.
In the observations, however, different photometric systems and transformation 
between them can cause difficult problems in 
comparing the extinction laws in different regions.  
Naoi et al. (2006) have shown that the photometry transformation 
between different photometric systems does not 
reproduce the real color excess ratio probably because the transformation is 
mainly determined from the photometry of intrinsically red stars, 
not reddened stars.  
Hence, Naoi et al. (2006) have conducted imaging observations for the 
$\rho$ Oph and Cha clouds with the same photometric system, and 
have not found any significant differences in the extinction law, \cer, 
between the clouds in contrast to previous works \citep{ken98,gom01}.
Naoi et al. (2006) have suspected that the apparent difference of 
the color excess ratio, \cer, 
was caused by the transformation between different photometric systems, 
at least for these two dark clouds.

Racca et al. (2002) observed the Coalsack and determined the 
color excess ratio $E_{J-H}/E_{H-K} = 2.08 \pm 0.03$,
which is significantly larger than the ratios for other clouds 
(the mean value of $1.61 \pm 0.04$, Whittet 1988).
They suggested possible changes in grain chemistry 
with extinction or star formation activity.  
Though the color excess ratios for 
the $\rho$ Oph ($1.57 \pm 0.03$, Kenyon et al. 1998), 
Cha ($1.80 \pm 0.03$, G\'omez \& Kenyon 2001), and 
Coalsack (Racca et al. 2002) clouds were all expressed in the same CIT system, 
the photometric system used in the actual observations are different, 
and the color transformation between them made the comparison very difficult. 

Hence we have conducted $J$, $H$, and $K_S$ imaging observations 
for the Coalsack Globule 2 (about 3 mag deeper than that of Racca et al. 2002) 
with the same photometric system (SIRIUS/IRSF) as in Naoi et al. (2006) 
for the $\rho$ Oph and Cha clouds, and 
determined the color excess ratio $E_{J-H}/E_{H-K_S}$. 
We discuss the extinction law of the Coalsack Globule 2 in addition to 
the $\rho$ Oph and Cha clouds \citep{nao06}, and 
also discuss the grain properties in comparison with 
extinction model calculations.


\section{Observations}

We acquired $J$ (1.25 $\mu$m), $H$ (1.63 $\mu$m), and $K_S$ (2.14 $\mu$m)
images of the Coalsack Globule 2 in March--April 2004 
with the Simultaneous Infrared Imager for Unbiased Survey (SIRIUS) on the
Infrared Survey Facility (IRSF) 1.4 m telescope at the South
African Astronomical Observatory (SAAO) in Sutherland, the Republic of 
South Africa. Dichroic mirrors enabled simultaneous observations
in the three bands (see Nagashima et al. 1999 and Nagayama et al. 2003 
for the details on the instrument). 
The camera is equipped with three 
1024$\times$1024 pixel HgCdTe (HAWAII) arrays. 
The pixel scale of the array was 0\farcs453 pixel$^{-1}$, 
giving a field of view of 7\farcm7$\times$7\farcm7.

We acquired 20 s exposure sets for 3 times. 
Ten dithered frames were obtained for one set of exposures. 
Total integration time is 10 minutes for each position. 
Relative positional offsets of dithering were 20\arcsec or 25\arcsec. 
Typical seeing conditions were 
1\farcs3, 1\farcs2, and 1\farcs1 (FWHM) 
in the $J$ , $H$, and $K_S$ bands, respectively. 
The observations were made at air masses between 1.0
and 1.5. Dark and twilight flat frames were obtained at the beginning
and end of each night. We observed standard stars from
the faint near-infrared standard-star catalog of Persson et al. (1998) 
for photometric calibration on the same night.

The target region 
encompasses heavily extinguished part 
of the Globule 2 in Coalsack determined from optical
extinction and radio CO maps 
\citep{rac02,kat99}.
We also observed a reference region in order to estimate the intrinsic color, 
which lies very close to the target region and
falls in regions of relatively low extinction \citep{rac02}.
The observed areas are summarized in Table \ref{tbl_obs}.

We used NOAO's Imaging Reduction and Analysis Facility
(IRAF)\footnote{IRAF is distributed by the 
National Optical Astronomy Observatory (NOAO), 
which is operated by the Association of Universities for Research in
Astronomy (AURA), Inc., under cooperative agreement with the National
Science Foundation.}
software package to reduce the data. We applied the
standard procedures of near-infrared array image reduction, including
dark current subtraction, sky subtraction, and flat fielding.
Each image, following subtraction of the average dark
frame, was divided by the normalized flat-field image. Then the
thermal emission pattern, the fringe pattern due to OH emission,
and the reset anomaly slope pattern of the HAWAII arrays were
corrected for each frame, with subtraction of a median sky
frame. Identification and photometry of point sources in the all
frames were performed with using the DAOPHOT packages in
IRAF. The $10\sigma$ limiting magnitudes ($\leq0.1$ mag) for point sources
were $J \sim 19.0$, $H\sim18.3$, and $K_S\sim17.0$ mag.
We used these point sources in analysis, 
but the point sources brighter than $J < 12.0$,
$H < 11.0$, and $K_S < 10.0$ (less than 1\% of the total) were not
included in our samples because they were saturated.


\section{Results and determining of the color excess ratio \cer}

\begin{figure}
\plottwo{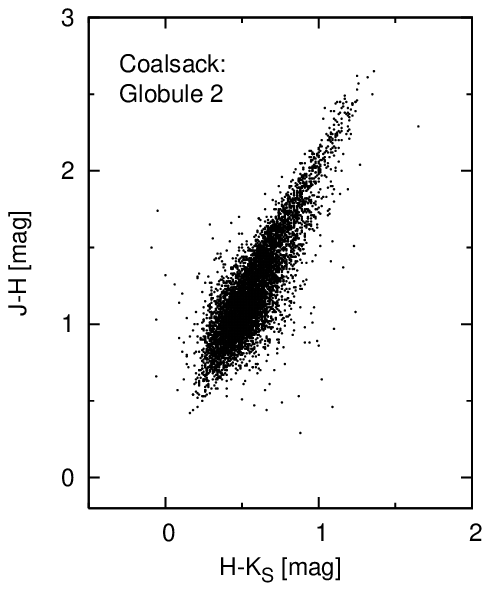}{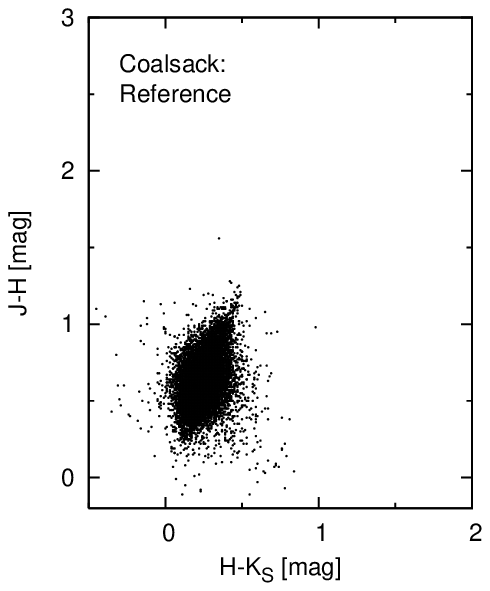}
\caption{$J-H$ vs. $H-K_S$ color-color diagrams of point sources 
for the Globule 2 and the reference region in Coalsack.\label{fig_dst}}
\end{figure}

Figure \ref{fig_dst} shows $J-H$ vs. $H-K_S$ color-color diagrams 
of point sources for the Globule 2 and the reference region in Coalsack. 
The photometry data described in this paper 
is in the SIRIUS/IRSF photometric system 
because we do not use any transformation equations to 
other photometric systems for photometry 
in order to avoid possible sources of errors \citep{nao06}.
In Table \ref{tbl_obs}, the numbers of identified point sources are also shown.

\begin{figure}
\plottwo{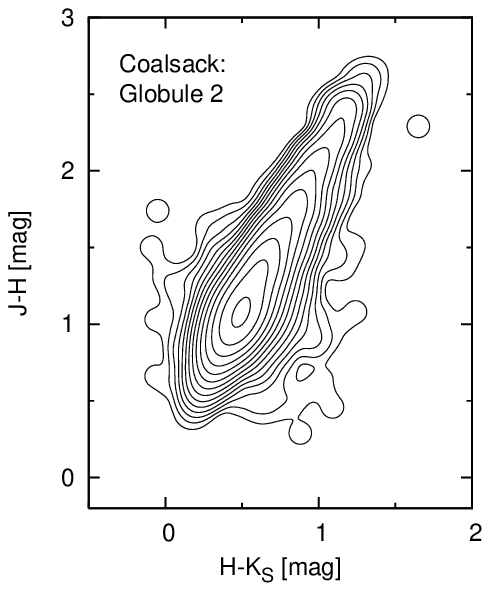}{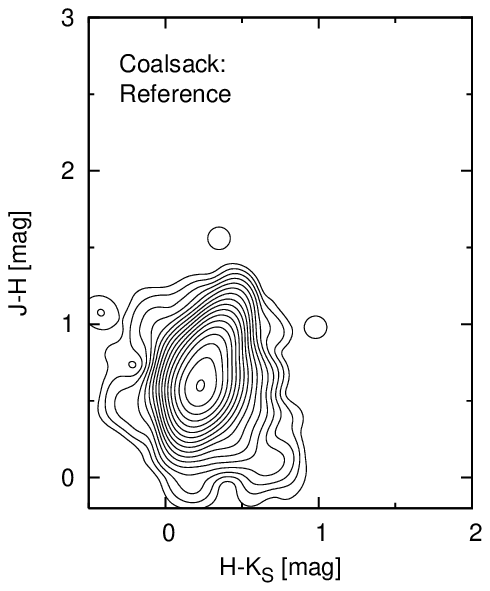}
\caption{$J-H$ vs. $H-K_S$ density distributions drawn from Fig.\ref{fig_dst}
for the Globule 2 and the reference region in Coalsack.\label{fig_ctr}}
\end{figure}

\begin{figure}
\plottwo{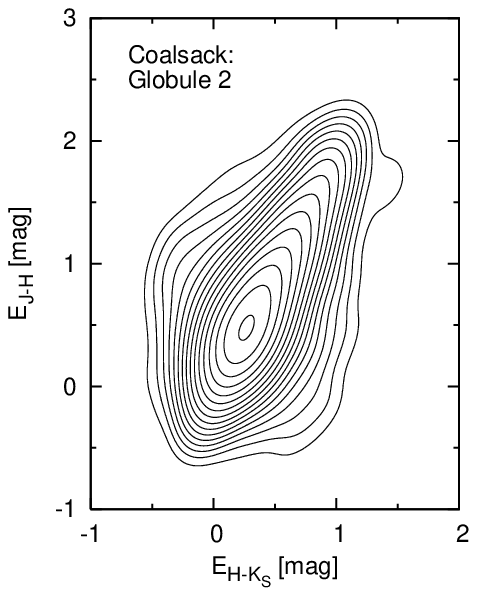}{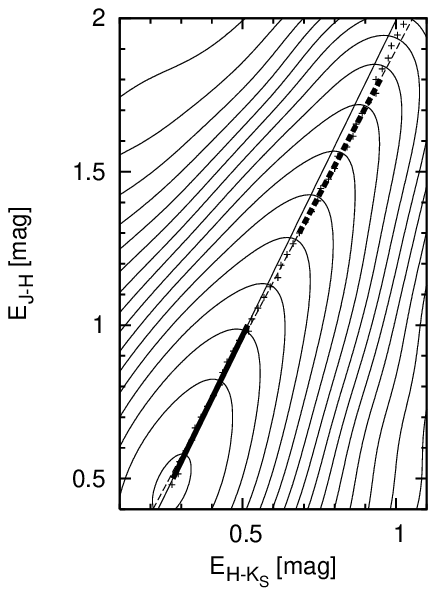}
\caption{$E_{J-H}$ vs. $E_{H-K_S}$ reddening probability contour 
for the Coalsack Globule 2 and the enlargement.
The plots of the ridgeline on contour (cross marks) and the best fit lines  
$E_{J-H} = 2.05 E_{H-K_S} - 0.06$ (for $0.5 < E_{J-H} < 1.0$; solid line) and 
$E_{J-H} = 1.89 E_{H-K_S} + 0.01$ (for $1.3 < E_{J-H} < 1.8$; dashed line)
are also shown in the enlargement.
The bold parts of the lines correspond to the fit ranges of the ridgeline.
\label{fig_prb}}
\end{figure}

In order to determine the color excess ratio, \cer, 
for the Coalsack Globule 2, we
employ the powerful method developed by Kenyon et al. (1998).
The method determines the ratio accurately even when a number of 
PMS stars are present in the sample \citep{nao06}. 
We assume
that the stellar population behind the cloud is identical to the stellar
population in the reference region. The point source distribution in a
color-color diagram is transformed to the density distribution by a
kernel density estimator with the kernel as triweight. 
Figure \ref{fig_ctr} shows 
$J-H$ vs. $H-K_S$ density distributions drawn from Figure \ref{fig_dst}
for the Globule 2 and the reference region in Coalsack.
Contours are drawn in a logarithmic scale
(the same as for all figures below). 
These color distributions use a kernel density estimator 
with a smoothing parameter $h = 0.2$ to derive the density, 
which is roughly twice the 1 $\sigma$ error 
of our photometry at the survey limits.
The sum of multiplications of the density distributions between the target and
reference regions provides the reddening probability distribution,
which is defined as the probability of measuring a pair of color
excesses, $E_{J-H}$ and $E_{H-K_S}$.
Figure \ref{fig_prb} shows $E_{J-H}$ vs. $E_{H-K_S}$ reddening 
probability contours for the Coalsack Globule 2.
We derive the slopes of the reddening vector, 
namely the color excess ratio $E_{J-H}/E_{H-K_S}$, 
by tracing the ridge of contours
on the reddening probability distribution which is divided
into the circular annuli centered on the origin. 
The slope of the reddening locus \cer is determined by a
least-squares fit of a straight line with data having errors in both
$x$ and $y$ coordinates (see Naoi et al. 2006 for more details).
The analysis yields 
$E_{J-H}/E_{H-K_S} = 1.91 \pm 0.01$ 
(for $0.50 < E_{J-H} < 1.8$) in the SIRIUS/IRSF photometric system, 
and the result is summarized in Table \ref{tbl_cer}.
In Figure \ref{fig_prb}, 
the enlargement with plots of the ridgeline on contour and the best fit lines 
for the different extinguished parts are also shown for clearness.


\section{Extinction law in the Coalsack Globule 2}

Table \ref{tbl_cer} shows 
the compilation of the color excess ratios, \cer,  
for the Coalsack Globule 2 and 
for the $\rho$ Oph and Cha clouds \citep{nao06}.
The color excess ratio for the Coalsack Globule 2 ($1.91 \pm 0.01$)
is significantly larger than those for the mean values of the 
$\rho$ Oph ($1.65 \pm 0.03$) and Cha ($1.67 \pm 0.01$) clouds.
Moreover, the value in the Coalsack Globule 2 is higher than those
in any sub-regions of the $\rho$ Oph and Cha clouds.
These results indicate that a real variation 
in the near-infrared extinction law occurs from cloud to cloud
(see also, Nishiyama et al. 2006 
for the case of the observations toward the Galactic Center).
The result in the Coalsack Globule 2 supports 
the relatively large ratio indicated by Racca et al. (2002). 
The shape of the extinction curve is frequently characterized 
by a law of the form $A_{(\lambda)} \propto \lambda^{-\alpha}$, 
where $\alpha$ is the spectral index for near-infrared extinction 
(e.g., Draine 2003).
A good fit to the curve for $J$, $H$, and $K_S$ bands is obtained with
$\alpha = 2.34 \pm 0.01$ by using the mean effective wavelength 
of the filters \citep{nis06}; 
this index is very large compared to the values in other regions 
(e.g., Whittet 1988 shows a mean value of $\alpha = 1.70 \pm 0.08$).

\begin{figure}
\plotone{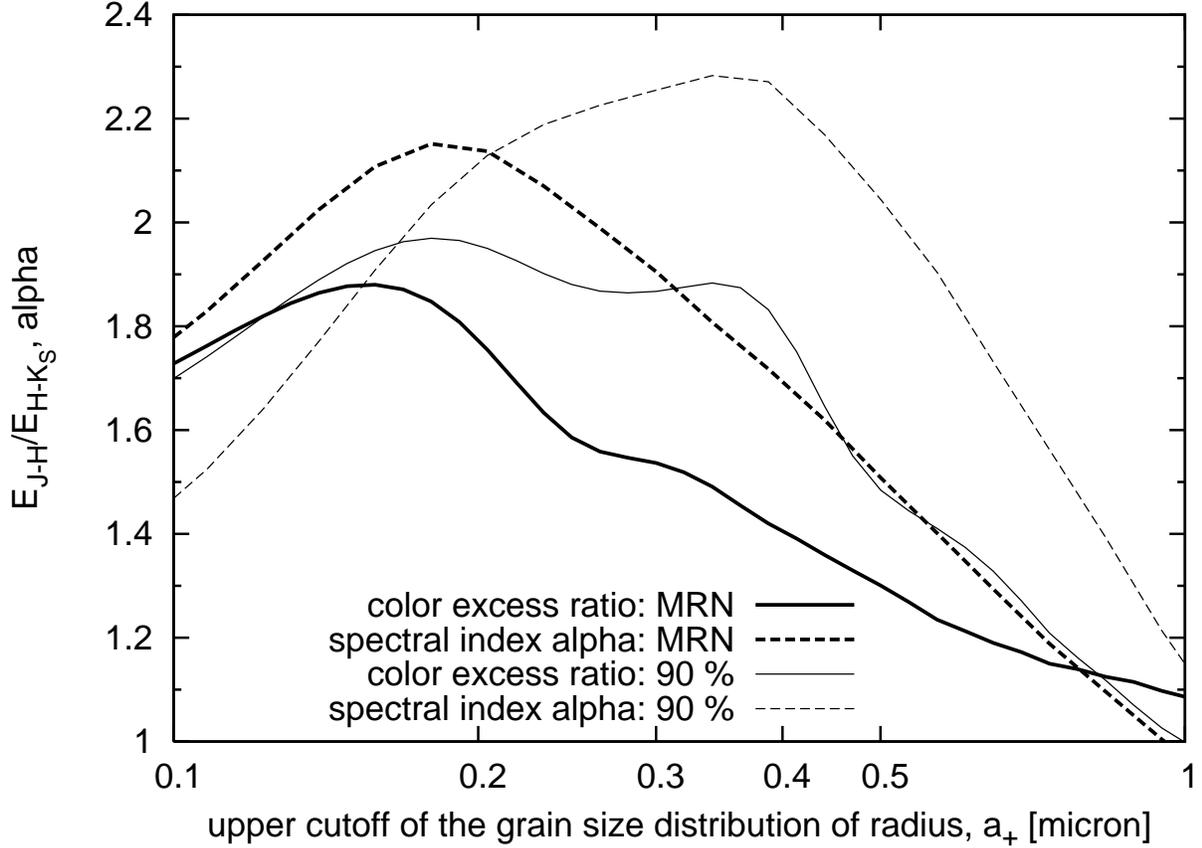}
\caption{Model calculations of the color excess ratio, $E_{J-H}/E_{H-K_S}$ 
(solid curves) and 
the spectral index, $\alpha$ (dashed curves) 
as a function of the upper cutoff of the grain size distribution of 
radius, $a_+$, 
for the different silicate/graphite ratios in MRN (53/47; bold curves) and
90 \% (90/10; fine curves). 
The large color excess ratio and spectral index in the Coalsack 
can be produced by increasing the abundance ratio of the silicate grain 
which is the representative material of non-absorbing grains. 
Spectral index $\alpha$ is fitted for the mean effective wavelengths of the 
$J$, $H$, and $K_S$ bands in the SIRIUS/IRSF photometric system \citep{nis06}.
\label{fig_calc}}
\end{figure}

Such a large color excess ratio, $E_{J-H}/E_{H-K_S}$, 
and a large spectral index, $\alpha$, in the Coalsack 
are realized when the upper cutoff of the grain size distribution of radius
is kept around 0.2 $\mu$m, 
and also if we assume greater dielectric material ratio.
Figure \ref{fig_calc} shows the calculations of the color excess ratio 
and the index $\alpha$ as a function of the upper cutoff of the 
grain size distribution of radius, $a_+$, and the silicate/graphite ratio 
in the simple MRN model in Naoi et al. (2006).
This may indicate that dust grains in the globule 
do not grow in size so much compared with the standard grain size 
$\sim 0.1$ $\mu$m in diffuse interstellar medium (e.g., Mathis 1977). 
Also, similar model calculations indicate that non-absorbing dielectric 
materials such as silicates cause rapidly decreasing extinction towards 
the longer wavelength, 
and result in large color excess ratios and large spectral indices
because the efficiency of the scattering cross section varies 
in inverse proportion to the fourth power of the wavelength 
(Rayleigh scattering, e.g., Bohren \& Huffman 1983).

We also confirm a systematic shift in the major axis 
of the probability contour in Figure \ref{fig_prb}. 
The plots of the ridgeline on contour and the best fit lines 
for the different extinguished parts 
are also shown in the enlargement.
The color excess ratio has a trend of decreasing from 
$E_{J-H}/E_{H-K_S} = 2.02$--2.08 for $0.5 < E_{J-H} < 1.0$ (solid line) to
$E_{J-H}/E_{H-K_S} = 1.87$--1.91 for $1.3 < E_{J-H} < 1.8$ (dashed line).
The trend is consistent with those in Naoi et al. (2006) for the 
$\rho$ Oph and Cha clouds.
Kenyon et al. (1998) suggests that the color excess ratio 
in the $\rho$ Oph cloud decreases from 
$E_{J-H}/E_{H-K} = 1.65$ for $0.3 < E_{J-H} < 0.4$ to
$E_{J-H}/E_{H-K} = 1.56$--1.57 for $1.5 < E_{J-H} < 1.8$.
Kenyon et al. (1998) also indicates a large color excess ratio 
$E_{J-H}/E_{H-K} \sim 2$ for the much weaker extinction range at 
$E_{J-H} < 0.2$ in the cloud. 
They suspect that this change might be statistically insignificant.  
In our case, however, the larger number of reference sources allows us 
to determine the ratio more precisely.
In order to verify the trend, the color excess ratios are determined 
for the two groups of the sample areas which are defined by the distance 
from the center of the Globule 2 in Kato et al. (1999); 
(a) the ``inside'' sample (radius $< 250$ arcsec), and
(b) the ``outside'' sample (radius $> 250$ arcsec). 
The results are complied in Table \ref{tbl_cer}.
The color excess ratio for the ``inside'' sample 
$E_{J-H}/E_{H-K_S} (\textrm{in}) = 2.01 \pm 0.03$ is 
smaller than that for the ``outside'' sample
$E_{J-H}/E_{H-K_S} (\textrm{out}) = 2.50 \pm 0.05$ 
in the same optical depth range of $0.7 < E_{J-H} < 1.2$.
The trend of the systematic shift in the major axis 
of the probability contour in Figure \ref{fig_prb} 
is thus verified.
As suggested by Naoi et al. (2006) for the $\rho$ Oph and Cha 
clouds, this indicates grain growth with optical depth, toward the 
inside of the dark cloud, also in the Coalsack Globule 2. 
Therefore, although the grain size in the Coalsack Globule 2 seems to be 
small compared with other dark clouds, a slight change in size might be 
taking place from its periphery to the central region.


\section{Conclusions}
\begin{enumerate}
\item
We have conducted $J$, $H$, and $K_S$ band imaging observations 
for the Coalsack Globule 2 with SIRIUS/IRSF 1.4m at SAAO and 
determined the color excess ratio, \cer,
by using the method in Kenyon et al. (1998) and Naoi et al. (2006).
\item
The color excess ratio for the Coalsack Globule 2 has a 
significantly larger value 
than those for the $\rho$ Oph and Cha clouds \citep{nao06}. 
Because of employing the same photometric system, 
our studies are simple and straightforward 
in comparison of near-infrared extinction laws.
\item
The extinction law for the Coalsack Globule 2 shows a trend that 
the color excess ratio decreases with 
increasing optical depth.
The trend indicated grain growth inside of the cloud, 
consistent with the results in Naoi et al. (2006) 
for the $\rho$ Oph and Cha clouds.
\item
The large color excess ratio for the Coalsack Globule 2 could be 
attributed to 
small grain growth and grain composition differences.
\end{enumerate}


\acknowledgments

We are grateful to the SIRIUS/IRSF and SAAO staff,
Yoshifusa Ita, Noriyuki Matsunaga, Yoshikazu Nakada, Toshihiko Tanab\'e, 
and Cynthia Strydom for their help in support of observations.
T.N. are also grateful to Koji Kawabata for the calculation of extinction, 
and University of Tokyo, NAOJ, and ISAS/JAXA staffs 
for support and assistance in this study. 
We wish to thank the anonymous referee and editors for careful comments.
This work is supported by grants-in-aid from 
the Japanese Ministry of Education, Science, and Culture 
(Nos. 12309010, 16340061, 16077204, and 15340061).




\clearpage

\clearpage

\begin{table}
\begin{center}
\caption{Center coordinates (J2000.0) of the target and reference regions 
and the number of identified point sources.\label{tbl_obs}}
\begin{tabular}{ccr}
\tableline\tableline
Region & R.A., Dec. & Number \\
\tableline
Globule 2 & 12 31 24, -63 45 11            & 7236 \\
          & (10\farcm0 $\times$ 10\farcm0) & \\ 
\tableline
Reference & 12 33 20, -64 24 39            & 15679 \\
          & (12\farcm0 $\times$ 12\farcm0) & \\
\tableline
\end{tabular}
\end{center}
\end{table}

\begin{table}
\begin{center}
\caption{Determined color excess ratios, \cer, for the Coalsack Globule 2 
with the $\rho$ Oph and Cha regions\tablenotemark{a} 
(SIRIUS/IRSF photometric system).\label{tbl_cer}}
\begin{tabular}{cccc}
\tableline\tableline
\multicolumn{2}{c}{Region} & \multicolumn{1}{c}{\cer} & 
\multicolumn{1}{c}{Range ($E_{J-H}$)} \\
\tableline
Coalsack   & Globule 2 & $1.91\pm0.01$ & 0.50--1.80 \\
           & inside\tablenotemark{b}    & $2.01\pm0.03$ & 0.70--1.20 \\
           & outside\tablenotemark{b}   & $2.50\pm0.05$ & 0.70--1.20 \\
$\rho$ Oph & L1688     & $1.60\pm0.01$ & 0.50--2.35 \\
           & L1689     & $1.66\pm0.01$ & 0.50--1.70 \\
           & L1712     & $1.68\pm0.01$ & 0.50--1.50 \\
       Cha & I         & $1.69\pm0.01$ & 0.50--2.00 \\
           & II        & $1.66\pm0.01$ & 0.50--1.80 \\
           & III       & $1.67\pm0.02$ & 0.50--1.30 \\
\tableline
\end{tabular}
\tablenotetext{a}{The color excess ratios \cer for the $\rho$ Oph and Cha 
clouds are from Naoi et al. (2006)}
\tablenotetext{b}{The samples of the ``inside'' group are selected by the 
distance (radius $< 250$ arcsec, the number of point sources is 3248) 
from the center coordinate in Kato et al. (1999). 
The samples of the ``outside'' are all the other point sources 
except for the ``inside'' samples (the number of point sources is 3988).}
\end{center}
\end{table}

\end{document}